\documentclass[12pt]{article}
\usepackage{epsf,latexsym}
\usepackage{amsmath,amsfonts,amssymb,amsthm,cite}

\theoremstyle{definition}                    

\theoremstyle{remark}

\numberwithin{equation}{section}             

\usepackage[ps,dvips,arc,frame]{xy}

\epsfverbosetrue
\newcommand{\mini}[1]{\stackrel{\circ}{#1}}

\newcommand{\ddf}{\hbox{$^f$\hspace{-0.15cm} $\mathcal{D}$}}

\newcommand{\ddfm}{\hbox{$^{\hat{f}}$\hspace{-0.15cm}
$\mathcal{D}$}}

\newcommand{\T}{{\rm tr}}

\newcommand{\bb}{\begin{eqnarray}}
\newcommand{\ee}{\end{eqnarray}}
\newcommand{\eee}{\nonumber\end{eqnarray}}
\newcommand{\pp}[1]{\begin{pmatrix} #1 \end{pmatrix}}

\newcommand{\rxyg}[2]{{\begin{xy} 0;<2mm,0mm>:<0mm,2mm>::0;0,
,(5,-2)*{a} ,(10,-1.8)*{b} ,(15,-2)*{c} ,(20,-1.8)*{d} ,(2,-5)*{a}
,(1.8,-10)*{b} ,(2,-15)*{c} ,(1.8,-20)*{d} ,(5,-5)*\cir(#1,0){}
,(10,-5)*\cir(#1,0){} ,(15,-5)*\cir(#1,0){} ,(20,-5)*\cir(#1,0){}
,(5,-10)*\cir(#1,0){} ,(10,-10)*\cir(#1,0){} ,(15,-10)*\cir(#1,0){}
,(20,-10)*\cir(#1,0){} ,(5,-15)*\cir(#1,0){} ,(10,-15)*\cir(#1,0){}
,(15,-15)*\cir(#1,0){} ,(20,-15)*\cir(#1,0){} ,(5,-20)*\cir(#1,0){}
,(10,-20)*\cir(#1,0){} ,(15,-20)*\cir(#1,0){} ,(20,-20)*\cir(#1,0){}
#2\end{xy}}}

\begin{document}

\thispagestyle{empty}

\begin{center}
CENTRE DE PHYSIQUE TH\'EORIQUE \footnote{\, Unit\'e Mixed de
Recherche (UMR) 6207 du CNRS et des Universit\'es Aix-Marseille 1 et 2 \\ \indent \quad \, Sud Toulon-Var, Laboratoire affili\'e \`a la 
FRUMAM (FR 2291)} \\ CNRS--Luminy, Case 907\\ 13288 Marseille Cedex 9\\
FRANCE
\end{center}

\vspace{1cm}

\begin{center}
{\Large\textbf{Massive Neutrinos in \\  Almost-Commutative
Geometry }} 
\end{center}

\vspace{1cm}

\begin{center}
{\large  Christoph A. Stephan$\,\,^{1,}$
\footnote{\, stephan@cpt.univ-mrs.fr} }

\vspace{1.5cm}

{\large\textbf{Abstract}}
\end{center}

In the noncommutative formulation of the standard model
of particle physics by A. Connes and A. Chamseddine \cite{cc} one
of the three generations of fermions has to possess  a 
massless neutrino. This formulation is consistent with
neutrino oscillation experiments and the known bounds
of the Pontecorvo-Maki-Nakagawa-Sakata matrix (PMNS matrix).
But future experiments which may be able to detect neutrino
masses directly and high-precission measurements of the
PMNS matrix might need massive neutrinos in all three
generations.

In this publication we present an almost-commutative geometry which allows
for a standard model with massive neutrinos in all three 
generations. This model does not follow in a straight forward way 
from Connes' and Chamseddine's version since it requires
an internal algebra with four summands of matrix algebras,
instead of three summands for the model with one massless
neutrino.

\vspace{1cm}

\vskip 1truecm

PACS-92: 11.15 Gauge field theories\\
\indent MSC-91: 81T13 Yang-Mills and other gauge theories

\vskip 1truecm

\noindent August 2006
\vskip 0.5truecm
\noindent \\
\noindent

\newpage

\section{Introduction}

Alain Connes' noncommutative geometry \cite{book,real,grav,cc} allows in an elegant way
to unify gravity and the standard model of particle physics. 
A central role in this formalism
is played by almost-commutative spectral triples ($\mathcal{A},\mathcal{H},\mathcal{D}$) which decompose into an external and an internal, finite dimensional component. The external
part encodes a compact 4-dimensional Euclidian spacetime and the internal one corresponds to 
a discrete 0-dimensional Kaluza-Klein space. 
The internal algebra consists of a sum of matrix algebras and  determines together
with the corresponding internal Hilbert space the particle content of the theory.
Via the spectral action  \cite{cc} one recovers the Einstein-Hilbert action combined with the bosonic
action of a Yang-Mills-Higgs (YMH) theory. The set of YMH theories compatible
with almost-commutative geometry is severely constrained. It could be shown that among the possible geometries those which produce the standard model of
particle physics take a most prominent position \cite{class}. The spectral triples were required to be irreducible and non-degenerate, in the sense that the 
Hilbert space was chosen to be as small as possible with non-degenerate fermion masses. For exact definitions see \cite{class}.
Heavy use was made of Krajewski's diagrammatic method \cite{Kraj}, which will be described briefly below. 

The standard
model with one generation of fermions turns out to be among the group 
of irreducible spectral triples with three and four summands in the internal algebra.
The case of three algebras
recovers the model of Connes and Chamseddine whereas the four algebra
case resembles more the Connes-Lott model \cite{cl}.
All of these minimal models do require a massless neutrino. It is possible to
add massive neutrinos if one extends the particle content to three generations,
but in the case of three summands at least one neutrino has to stay massless.
In this paper we will present a model with massive neutrinos in all three 
generations based on the irreducible version of the standard model with 
four summands in the internal algebra.  

\section{Basic Definitions}

In this section we will give the necessary basic definitions  of almost commutative
geometries from a particle physics point of view. 
For our calculations, only the finite part matters, 
so we restrict ourselves to real, $S^0$-real, finite spectral triples
($\mathcal{A},\mathcal{H},\mathcal{D}, $ $J,\epsilon,\chi$). The algebra $\mathcal{A}$ is
a finite sum of matrix algebras
$\mathcal{A}= \oplus_{i=1}^{N} M_{n_i}(\mathbb{K}_i)$ with $\mathbb{K}_i=\mathbb{R},\mathbb{C},\mathbb{H}$ where $\mathbb{H}$
denotes the quaternions. 
A faithful representation $\rho$ of $\mathcal{A}$ is given on the finite dimensional Hilbert space $\mathcal{H}$.
The Dirac operator $\mathcal{D}$ is a selfadjoint operator on $\mathcal{H}$ and plays the role of the fermionic mass matrix.
$J$ is an antiunitary involution, $J^2=1$, and is interpreted as the charge conjugation
operator of particle physics.
The $S^0$-real structure $\epsilon$ is a unitary involution, $\epsilon^2=1$. Its eigenstates with
eigenvalue $+1$ are the particle states, eigenvalue $-1$ indicates antiparticle states. 
The chirality $\chi$ as well is a unitary involution, $\chi^2=1$, whose eigenstates with eigenvalue
$+1$ $(-1)$ are interpreted as right (left) particle states.
These operators are required to fulfill Connes' axioms for spectral triples:

\begin{itemize}
\item  $[J,\mathcal{D}]=[J,\chi]=[\epsilon,\chi]=[\epsilon,\mathcal{D}]=0, \quad \epsilon
J=-J \epsilon,\quad\mathcal{D}\chi =-\chi \mathcal{D}$, 
 
$[\chi,\rho(a)]=[\epsilon,\rho(a)]=[\rho(a),J\rho(b)J^{-1}]=
[[\mathcal{D},\rho(a)],J\rho(b)J^{-1}]=0, \forall a,b \in \mathcal{A}$.
\item The chirality can be written as a finite sum $\chi =\sum_i\rho(a_i)J\rho(b_i)J^{-1}.$
This condition is called {\it orientability}.
\item The intersection form
$\cap_{ij}:=\T(\chi \,\rho (p_i) J \rho (p_j) J^{-1})$ is non-degenerate,
$\rm{det}\,\cap\not=0$. The
$p_i$ are minimal rank projections in $\mathcal{A}$. This condition is called
{\it Poincar\'e duality}.
\end{itemize} 
Now the Hilbert space $\mathcal{H}$ and the representation $\rho$ decompose with respect to the 
eigenvalues of $\epsilon$ and $\chi$ into left and right, particle and antiparticle spinors
and representations:
\begin{eqnarray}
\mathcal{H}=\mathcal{H}_L\oplus\mathcal{H}_R\oplus\mathcal{H}_L^c\oplus\mathcal{H}_R^c \quad \quad 
\rho = \rho_L \oplus \rho_R \oplus \overline{ \rho_L^c} \oplus \overline{ \rho_R^c}
\notag
\label{representation}
\end{eqnarray}
In this representation the Dirac operator has the form
\begin{eqnarray}
\mathcal{D}=\pp{0&\mathcal{M}&0&0\\
\mathcal{M}^*&0&0&0\\ 0&0&0&\overline{\mathcal{M}}\\
0&0&\overline{\mathcal{M}^*}&0}, \label{opdirac}
\notag
\end{eqnarray}
where $\mathcal{M}$ is the fermionic mass matrix connecting the left and the right handed fermions.

Connes' axioms, the decomposition of the Hilbert space, the representation and the Dirac operator
allow a diagrammatic depiction, known as Krajewski diagrams. 
As was shown in \cite{Kraj} and \cite{class} this can be boiled down
to simple arrows, which encode the intersection form and the fermionic mass matrix.
From these informations all the ingredients of the spectral triple can be recovered.
For our purpose a simple arrow and a double arrow are sufficient. The arrows allways point from
right fermions (positive chirality) to left fermions (negative chirality).
 We may also restrict ourselves to
the particle part, since the information of the antiparticle part is included by transposing the
particle part.We will adopt
the conventions of \cite{class}.

To complete our short survey, we will give a brief glimpse
on how to construct the actual Yang-Mills-Higgs theory.
We started out with the fixed (for convenience flat) Dirac operator of a 4-dimensional spacetime with a fixed fermionic
mass matrix. To generate curvature we have to perform a general coordinate transformation and then
fluctuate the Dirac operator. This can be achieved by lifting the automorphisms of the algebra to
the Hilbert space, unitarily transforming the Dirac operator with the lifted automorphisms
and then building linear combinations. Again we restrict ourselves to the finite case.
Except for complex conjugation in $M_n(\mathbb{C})$ and permutations of
identical summands in the algebra $\mathcal{A}=\mathcal{A}_1\oplus\mathcal{A}_2\oplus ...\oplus\mathcal{A}_N$,
every algebra automorphism
$\sigma
$  is inner, $\sigma (a)=uau^{-1}$ for a unitary $ u\in U(\mathcal{A})$. Therefore
the connected component of the automorphism group is
Aut$(\mathcal{A})^e=U(\mathcal{A})/(U(\mathcal{A})\cap{\rm Center}(\mathcal{A}))$. Its lift to the Hilbert
space \cite{real}
\bb 
L(\sigma )=\rho (u)J\rho (u)J^{-1}
\label{centralextendlift}
\ee 
is multi-valued. To avoid the multi-valuedness in the fluctuations, we allow  a central extension of the automorphism group. As we will see, central extensions will also allow to cancel anomalies.

The {\it fluctuation $\ddf$} of the Dirac operator $\mathcal{D}$ is given by a
finite collection $f$ of real numbers
$r_j$ and algebra automorphisms $\sigma _j\in{\rm Aut}(\mathcal{A})^e$ such
that
\bb
\ddf :=\sum_j r_j\,L(\sigma _j) \, \mathcal{D} \, L(\sigma_j)^{-1},\quad r_j\in\mathbb{R},\
\sigma _j\in{\rm Aut}(\mathcal{A})^e.
\eee
We consider only fluctuations with real coefficients since $\ddf$ must remain selfadjoint.
The sub-matrix of the fluctuated Dirac operator $\ddf$ which is equivalent to
the mass matrix $\mathcal{M}$,  is often denoted by $\varphi $, the
`Higgs scalar', in physics literature.

An almost commutative geometry is the tensor product of a finite
noncommutative triple with an infinite, commutative spectral triple. By
Connes' reconstruction theorem \cite{grav} we know that the latter comes
from a Riemannian spin manifold, which we will take to be any
4-dimensional, compact, flat manifold like the flat 4-torus.  The spectral
action of this almost commutative spectral triple reduced to the finite part
is a functional on the vector space of all fluctuated, finite Dirac operators:
\bb V(\ddf )= \lambda\  \T\!\left[ (\ddf )^4\right] -\frac{\mu
^2}{2}\
\T\!\left[
(\ddf) ^2\right] ,\eee where $\lambda $ and $\mu $ are positive constants
\cite{cc}.
The spectral action is invariant under lifted automorphisms and even
under the unitary group $U(\mathcal{A})\owns u$,
\bb V( [\rho (u)J\rho (u)J^{-1}] \, \ddf \, [\rho (u)J\rho
(u)J^{-1}]^{-1})=V(\ddf),\eee and it is bounded from below.
To obtain the physical content of a diagram and its associated spectral triple one has
to find the minima $\ddfm $ of this action

\section{The Model}

We will start with the Krajewski diagram of the standard model with four summands
in the matrix algebra and only one fermion generation as it was shown in \cite{class}:
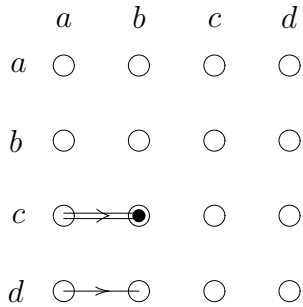
\begin{center}
\begin{tabular}{c}
\rxyg{0.7}{
,(10,-15)*\cir(0.4,0){}*\frm{*}
,(5,-20);(10,-20)**\dir{-}?(.6)*\dir{>}
,(5,-15);(10,-15)**\dir2{-}?(.6)*\dir2{>}
} \\ \\
\end{tabular}
\end{center}
Adding a massive neutrino, which also means adding a right-handed neutrino,
can be achieved by drawing an extra arrow.
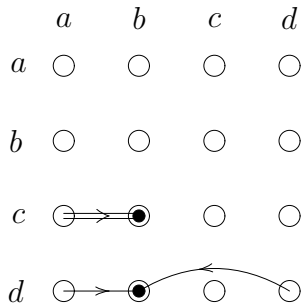
\begin{center}
\begin{tabular}{c}
\rxyg{0.7}{
,(10,-15)*\cir(0.4,0){}*\frm{*}
,(10,-20)*\cir(0.4,0){}*\frm{*}
,(5,-20);(10,-20)**\dir{-}?(.6)*\dir{>}
,(20,-20);(10,-20)**\crv{(15,-17)}?(.6)*\dir{>}
,(5,-15);(10,-15)**\dir2{-}?(.6)*\dir2{>}
}
\end{tabular}
\bb \label{neut} \ee
\end{center}
The determinant of the corresponding multiplicity matrix
\bb
\mu = \pp{ 0 & 0 & 2 & 1 \\ 0&0& -1 & -1 \\ 2 & -1 & 0 & 0 \\ 1 & -1 & 0 & 2},
\ee
is non-zero and so the axioms of noncommutative geometry remain fulfilled.
Note that adding the right-handed neutrino by copying the form of the quark sector
does not work,
\begin{center}
\begin{tabular}{c}
\rxyg{0.7}{
,(10,-15)*\cir(0.4,0){}*\frm{*}
,(10,-20)*\cir(0.4,0){}*\frm{*}
,(5,-20);(10,-20)**\dir2{-}?(.6)*\dir2{>}
,(5,-15);(10,-15)**\dir2{-}?(.6)*\dir2{>}
} \\ \\
\end{tabular}
\end{center}
since the determinant of the multiplicity matrix
\bb
\mu = \pp{ 0 & 0 & 2 & 1 \\ 0&0& -1 & -1 \\ 2 & -1 & 0 & 0 \\ 2 & -1 & 0 & 0},
\ee
is zero. Due to this reason it is impossible to add a right-handed neutrino
in the case of three summands in the internal algebra.

Now the internal algebra is chosen to be  $\mathcal{A} = \mathbb{C} \oplus \mathbb{H} \oplus M_3(\mathbb{C}) \oplus \mathbb{C} \ni (a,b,c,d)$
From the Krajewski diagram \ref{neut} one reads off the representation of the algebra, ordered into left and right, particle and antiparticle part:
\bb
\rho _L (a,b,c,d) &=&\pp{b\otimes 1_3&0\\ 0& b},\quad
\rho _R (a,b,c,d) =\pp{a 1_3&0&0&0\\ 0& \bar{a} 1_3&0&0\\ 0&0& \bar{a}&0 \\ 0&0&0&\bar{d}},
\nonumber \\
\nonumber \\
\rho _L^c (a,b,c,d) &=&\pp{1_2\otimes c&0\\ 0& d 1_2},\quad
\rho _R^c (a,b,c,d) =\pp{c&0&0&0\\ 0&c&0&0\\ 0&0& d& 0 \\ 0&0&0&d},
\ee
where $1_3$ is the unit matrix and the complex conjugates where chosen
in order to reproduce the standard model. This representation acts on the Hilbert
space $\mathcal{H} = \mathcal{H}^{PL} \oplus \mathcal{H}^{PR} \oplus \mathcal{H}^{AL} \oplus \mathcal{H}^{AR}$
with the left-  and right-handed particle subspaces 
\bb
\mathcal{H}^{PL} = \pp{\pp{u \\ d}^L \\ \pp{ \nu_e \\ e}^L}, \quad 
\mathcal{H}^{PR} = \pp{ u^R \\ d^R \\ \nu_e^R \\e^R } 
\ee
and antiparticle subspaces
\bb
\mathcal{H}^{AL} = \pp{\pp{u^c \\ d^c}^L \\ \pp{ \nu_e^c \\ e^c}^L}, \quad
\mathcal{H}^{AR} = \pp{ u^{cR} \\ d^{cR} \\ \nu_e^{cR} \\ e^{cR} }.
\ee
Note the fermions automatically appear as left-handed
doublets and right-handed singlets. 

The particle part $\Delta$ of the Dirac operator is
\bb
\Delta = \pp{0&0& M_1 \otimes 1_3 & M_2 \otimes 1_3 &0&0 \\
0 &0&0&0& M_3 & M_4\\
M_1^{\ast} \otimes 1_3 &0&0&0&0 &0\\
M_2^{\ast} \otimes 1_3 &0&0&0&0&0 \\
0&M_3^{\ast} &0&0&0&0 \\
0&M_4^{\ast} &0&0&0&0}
\ee
We choose the four mass matrices $M_1$, $M_2$, $M_3$
and $M_4$ of the initial Dirac operator  as
\bb
M_1 = \pp{m_1 \\ 0},\; M_2 = \pp{0 \\ m_2},\; \; M_3 = \pp{m_3 \\ 0},\; \mbox{and} \; M_4 = \pp{0 \\ m_4},
\label{massmatrices}
\ee
with $m_1,m_2,m_3,m_4 \in \mathbb{C}$ arbitrary.

In the next step the lift $L$ has to be worked out. The unitaries of
$\mathcal{A}$ close to the identity are: $\mathcal{U}^e(\mathbb{C})=U(1)$,   
$\mathcal{U}^e(\mathbb{H})=SU(2)$ and $\mathcal{U}^e(M_3(\mathbb{C}))=U(3)$.
With definition \ref{centralextendlift}
and the algebra representation one finds for the particle part of the lift
\bb
L^P((\det w)^p ,u,(\det w)^q w, (\det w)^r) &=& \rho^P_L (\cdots )
\rho^A_L (\cdots)  \oplus \rho^P_R (\cdots)
 \rho^A_R(\cdots)
\nonumber \\
&=& {\rm diag}\left[(\det w)^q u\otimes w, (\det w)^r u,  \right. \nonumber
 \\ 
&& \quad \quad  (\det w)^{p+q} w, (\det w)^{q-p} w, 
\\
&& \quad \quad \left. (\det w)^0, (\det w)^{-p + r}\right],
\nonumber 
\label{Standlift}
\ee
where $u\in SU(2)$, $w \in U(3)$ and $p,q,r \in \mathbb{Q}$. The exponents $p$, $q$ and $r$ are fixed to recover the standard model charges up to normalisation
\bb
q = \frac{p-1}{3} \quad {\rm and} \quad r = -p.
\ee
Next the Higgs potential $\varphi$ can be calculated. It follows for the particle part of the Higgs potential
\bb
\varphi^P &=& \sum_i r_i L^P((\det w_i)^p ,u_i,(\det w_i)^{\frac{p}{3}} \tilde{w_i},
(\det w_i)^{-p})  \times \Delta \nonumber
\\ 
&& 
\quad \times \left[L^P((\det w_i)^p ,u_i,(\det w_i)^{\frac{p}{3}} \tilde{w_i},(\det w_i)^{-p} )\right]^{-1}
\nonumber \\ \nonumber \\
&=& \pp{0&0&\varphi_1 \otimes 1_3 & \varphi_2 \otimes 1_3 &0 &0 \\
0&0&0&0& \varphi_3  & \varphi_4 \\
\varphi_1^{\ast} \otimes 1_3 &0&0&0&0 &0\\
\varphi_2^{\ast} \otimes 1_3 &0&0&0&0 &0\\
0 &\varphi_3^{\ast}&0&0&0&0 \\
0 &\varphi_4^{\ast}&0&0&0&0}
\ee
with $\tilde w= (\det w)^{-1/3} w \in SU(3)$ and
\bb
\varphi_{1} &=& \sum_i r_i u_i M_{1} ({\rm det}w_i)^{-p}, \;  \;\varphi_{2} = \sum_i r_i u_i M_{1} ({\rm det}w_i)^{p},
\nonumber \\
\varphi_{3} &=& \sum_i r_i u_i M_{3} ({\rm det}w_i)^{-p}, \;  \;\varphi_{4} = \sum_i r_i u_i M_{4} ({\rm det}w_i)^{p}.
\ee
For the Dirac operator with mass matrices \ref{massmatrices} it follows that $\varphi_1$
and $\varphi_2$ as well as  $\varphi_3$ and $\varphi_4$
allow the following operations: One can combine  $\varphi_1$ and $\varphi_2$ 
($\varphi_3$ and $\varphi_4$) into a single
matrix
\bb
\varphi_{m/n} &=& \sum_i r_i u_i (M_{m},M_{n}) \pp{({\rm det}w_i)^{-p} & 0 \\  0 &({\rm det}w_i)^{p}}
\nonumber \\ \nonumber \\
&=& \sum_i r_i u_i \pp{({\rm det}w_i)^{-p} & 0 \\  0 &({\rm det}w_i)^{p}}(M_{m},M_{n}).
\ee
This is possible since $(M_m,M_n)$ are diagonal matrices. One sees further that 
\bb
\pp{({\rm det}w_i)^{-p} & 0 \\  0 &({\rm det}w_i)^{p}} \in SU(2)
\ee
and thus
\bb
u_i \pp{({\rm det}w_i)^{-p} & 0 \\  0 &({\rm det}w_i)^{p}} \in SU(2).
\ee
With the observation that any quaterion $h\in\mathbb{H}$ can be written as $h = r u$ with
$r \in  \mathbb{R}$ and $u \in SU(2)$ it follows that $\varphi_{1/2}$ and $\varphi_3/4$
take the simple form
\bb
\varphi_{m/n} = h \pp{m_m &0 \\ 0& m_n} \; \; \mbox{with} \;
\; h = \pp{x & y \\ -\bar{y} & \bar{x}} \in \mathbb{H}.
\ee
It is now easy to minimise the Higgs potential
\bb
V(\varphi) = \lambda \mbox{tr} (\varphi^{\ast} \varphi)^2 - \frac{1}{2} \mu^2 \mbox{tr} (\varphi^{\ast} \varphi)
\ee
with respect to the variables $x,y \in \mathbb{C}$. To break up  the calculation into smaller steps one starts with
\bb
\varphi^{\ast} \varphi = \pp{ \varphi_{1/2} \varphi_{1/2}^{\ast} \otimes 1_3 &0&0&0 \\
0& \varphi_{3/4} \varphi_{3/4}^{\ast}&0&0 \\
0&0& \varphi_{1/2}^{\ast} \varphi_{1/2} \otimes 1_3 & 0 \\
0& 0&0& \varphi_{3/4}^{\ast} \varphi_{3/4} }.
\ee
It follows that 
\bb
\mbox{tr}  (\varphi \varphi^{\ast}) &=& 6\, \mbox{tr} (\varphi_{1/2}^{\ast} \varphi_{1/2}) + 2 \, \mbox{tr}(\varphi_{3/4} \varphi_{3/4}^{\ast}) 
\ee
and
\bb
\mbox{tr} ( \varphi \varphi^{\ast})^2 &=& 6 \, \mbox{tr} ( \varphi_{1/2}^{\ast} \varphi_{1/2})^2 + 2 \, \mbox{tr}(\varphi_{3/4} \varphi_{3/4}^{\ast})^2. 
\ee
Furthermore one has
\bb
\varphi_{m/n}^{\ast} = \pp{\bar{m}_m  & 0 \\ 0 & \bar{m}_n } \pp{\bar{x} & -y \\ \bar{y} & x}
\ee
and 
\bb
\varphi_{m/n}^{\ast}\varphi_{m/n} &=& \pp{\bar{m}_m  & 0 \\ 0 & \bar{m}_n } \pp{\bar{x} & -y \\ \bar{y} & x}
\pp{x & y \\ -\bar{y} & \bar{x}} \pp{m_m &0 \\ 0& m_n} \nonumber \\ \nonumber \\
&=& \pp{\bar{m}_m  & 0 \\ 0 & \bar{m}_n } \pp{|x|^2 + |y|^2 &0 \\ 0 &|x|^2 + |y|^2} \pp{m_m &0 \\ 0& m_n}
\nonumber \\ \nonumber \\
&=& r \pp{|m_m|^2  & 0 \\ 0 & |m_n|^2 }, 
\ee
with $r := |x|^2 + |y|^2$ as the new variable. 
Putting everything into $V(\varphi)$ gives
\bb
V(\varphi) &=& \lambda ( 6\, |m_1|^4 + 6\, |m_2|^4 + 2 \, |m_3|^4 + 2\, |m_4|^4) \, r^2 \nonumber \\
&&- \frac{\mu^2}{2} ( 6\, |m_1|^2 + 6\, |m_2|^2 + 2 \, |m_3|^2 + 2\, |m_4|^2 ) \, r 
\nonumber \\
&=:& \lambda \alpha  \, r^2 - \frac{\mu^2}{2} \beta  \, r.
\ee
The minimum in $r$ is found by differentiating with respect to $r$ and is 
\bb
\mini{r}\, = \frac{\alpha}{\beta} \frac{\mu^2}{4 \lambda}.
\ee
Now one can construct a minimum of the Higgs potential as
\bb
\mini{\varphi} = \sqrt{\mini{r}} \, \Delta.
\ee
The minimum is given by the initial mass
matrices $M_1$, $M_2$, $M_3$ and $M_4$ up to a fixed numerical factor $\sqrt{\mini{r}} \in \mathbb{R}^+$.
The masses $m_1$, $m_2$, $m_3$ and $m_4$ are identified as the quark masses, the neutrino mass and electron mass.

In the last step the little group and the charges of the fermions have to be found. 
$U(1)\times SU(2) \times SU(3)$ is the unbroken gauge group. 
The little group $G_\ell$ is defined by $\rho^P_L(g_\ell)\mini{\varphi}-\mini{\varphi}\rho^P_R(g_\ell) =0$ for all $g_\ell \in G_\ell$.
This is only possible if the representation is diagonal.
It follows that $U(1) \times SU(2) \rightarrow U(1)\subset SU(2) \ni v$  so that $G_\ell = U(1) \times SU(3)$.
The charges of the fermions are then given by the lift of $G_\ell$:
\bb
L^{PL}  &=&
\pp{ (\det w)^{\frac{p}{3}} \pp{v &0 \\ 0 & \bar{v}} \otimes \tilde{w} &0 \\
0& (\det w)^p \pp{v &0 \\ 0 & \bar{v}} } 
\nonumber \\ \nonumber \\ 
L^{PR}  &=&\pp{ 
 (\det w)^{p + \frac{p}{3}} \tilde{w}&0&0&0 \\
0  &(\det w)^{-p + \frac{p}{3}} \tilde{w} &0 &0\\
0&0 &(\det w)^{0}&0 \\
0&0&0 &(\det w)^{-2p} } ,
\ee
the electric charges being the exponents of the $U(1)$ elements.

Finally $p$ and $v$ have to be fixed in view of experimental data. The neutrino has to be
neutral under the little group and so $(\det w)^p v =1$ and it follows that 
$v = (\det w)^{-p}$.
With $p=-1/2$ the electric charge of the electron is $Q_e=-1$ and the quark charges are
$Q_d=-1/3$ and  $Q_u=+2/3$. Note that the charges for the left-handed and
right-handed fermions are equal and that the colour and the electric charge couple vectorially, as desired. The right-handed neutrino is completely neutral.
We also remark at this point that the gauge and mixed gravitational anomalies can be canceled only by the use of a particular central extension.

\section{Conclusion} 

We presented an almost-commutative geometry with four summands in
the internal algebra which allows massive neutrinos in all generations of
the standard model. This model is reducible in the sense of \cite{class},
i.e. the right-handed neutrino can be erased without violating any
axiom of noncommutative geometry. One may speculate that this
reducibility is connected to the relatively small mass of the neutrino
compared with the other fermions.

It should be pointed out that from the experimental point of view a massless
neutrino in one of the fermion generations is not excluded. For a 
complete classification of all mass matrices and  corresponding PMNS
mixing matrices compatible with the experiment we refer to \cite{hage}.

\vskip1cm
\noindent
{\bf Acknowledgements:} The author would like to thank T. Sch\"ucker for careful proof
reading. We
gratefully acknowledge a fellowship of the Alexander von Humboldt-Stiftung.


\begin{thebibliography}{10}

\bibitem{cc}
 A. Chamseddine \& A. Connes, ``The spectral action principle'',
hep-th/9606001, Comm. Math. Phys. 182 (1996) 155
\bibitem{book}
 A. Connes, {\it Noncommutative Geometry}, Academic Press, London and San
Diego (1994)
\bibitem{real}
A. Connes, ``Noncommutative geometry and reality'',  J.
Math. Phys. 36 (1995) 6194
\bibitem{grav}
A. Connes, `` Gravity coupled with matter and the
foundation of noncommutative geometry'', hep-th/9603053, Comm. Math.
Phys. 155 (1996) 109
\bibitem{class}
B. Iochum, T. Sch\"ucker, C. Stephan,
``On a Classification of Irreducible Almost Commutative Geometries'',
hep-th/0312276, J.Math.Phys. 45 (2004) 5003
\\
J.-H. Jureit, C. Stephan,
``On a Classification of Irreducible Almost Commutative Geometries, a Second Helping'',
hep-th/0501134, J.Math.Phys. 46 (2005) 043512
\\
J.-H. Jureit, T. Sch\"ucker, C. Stephan,
``On a Classification of Irreducible Almost Commutative Geometries III'',  hep-th/0503190, J.Math.Phys. 46 (2005) 072303
\\
T. Sch\"ucker
``Krajewski diagrams and spin lifts'',
hep-th/0501181
\bibitem{Kraj}
T. Krajewski, ``Classification of finite spectral triples'',
hep-th/9701081, J. Geom. Phys. 28 (1998) 1
\bibitem{cl}
A. Connes \& J. Lott, ``Particle models and noncommutative
geometry'', Nucl. Phys. B 18B (1990) 29\\
 A. Connes \& J. Lott,``The metric
aspect of noncommutative geometry'', in the
proceedings of the 1991 Carg\`ese Summer Conference,
eds.: J. Fr\"ohlich et al., Plenum Press (1992)
\bibitem{hage}
C. Hagedorn \& W.Rodejohann, ``Minimal mass matrices for Dirac neutrinos'',
hep-ph/0503143
\end{thebibliography}
\end{document}